%% file: HMXBs_FormEff.tex
\documentclass[twocolumn,floatfix]{aastex62}

\hypersetup{linkcolor=red,citecolor=green,filecolor=cyan,urlcolor=magenta}
\submitjournal{ApJ}

\shorttitle{Formation efficiency of High-Mass X-ray Binaries}
\shortauthors{Antoniou et al.}
\usepackage{color}

\include{definitions}

\begin{document}

\title{Deep \chandra survey of the Small Magellanic Cloud. III. Formation efficiency of high-mass X-ray binaries}

\correspondingauthor{Vallia Antoniou}
\email{vantoniou@cfa.harvard.edu}

\author[0000-0001-7539-1593]{Vallia Antoniou}
\affil{Harvard-Smithsonian Center for Astrophysics 
 60 Garden Street, Cambridge, MA 02138, USA}
\affil{Department of Physics and Astronomy, Box 41051, Science Building, Texas Tech University, Lubbock, TX 79409, USA}

\author{Andreas Zezas}
\affiliation{Physics Department and Institute of Theoretical and Computational Physics, University of Crete, 71003 Heraklion, Crete, Greece; Foundation for Research and Technology-Hellas, 71110 Heraklion, Crete, Greece}
\affiliation{Harvard-Smithsonian Center for Astrophysics 
 60 Garden Street, Cambridge, MA 02138, USA}

\author{Jeremy J. Drake}
\affiliation{Harvard-Smithsonian Center for Astrophysics 
 60 Garden Street, Cambridge, MA 02138, USA}

\author{Carles Badenes}
\affiliation{Department of Physics and Astronomy and Pittsburgh Particle Physics, Astrophysics and Cosmology Center, University of Pittsburgh, Pittsburgh, PA 15260, USA}

\author{Frank Haberl}
\affiliation{Max-Planck-Institut f\"{u}r extraterrestrische Physik, Giessenbachstraße, D-85748 Garching, Germany}

\author{Nicholas J. Wright}
\affiliation{Astrophysics Group, Keele University, Keele, ST5 5BG, UK}

\author{Jaesub Hong}
\affiliation{Harvard-Smithsonian Center for Astrophysics 
 60 Garden Street, Cambridge, MA 02138, USA}

\author{Rosanne Di Stefano}
\affiliation{Harvard-Smithsonian Center for Astrophysics 
 60 Garden Street, Cambridge, MA 02138, USA}

\author{Terrance J. Gaetz}
\affiliation{Harvard-Smithsonian Center for Astrophysics 
 60 Garden Street, Cambridge, MA 02138, USA}

\author{Knox S. Long}
\affiliation{Space Telescope Science Institute, 3700 San Martin Drive, Baltimore, MD 21218, USA}

\author{Paul P. Plucinsky}
\affiliation{Harvard-Smithsonian Center for Astrophysics 
 60 Garden Street, Cambridge, MA 02138, USA}

\author{Manami Sasaki}
\affiliation{Dr. Karl Remeis-Sternwarte, Erlangen Centre for Astroparticle Physics, Friedrich-Alexander-Universität Erlangen-Nürnberg, Sternwartstrasse 7, D-96049 Bamberg, Germany}

\author{Benjamin F. Williams}
\affiliation{Astronomy Department, University of Washington, Box 351580, Seattle, WA 98195, USA}

\author{P. Frank Winkler}
\affiliation{Department of Physics, Middlebury College, Middlebury, VT 05753, USA}
 
\collaboration{(SMC XVP collaboration)}

%% Note that the \and command from previous versions of AASTeX is now
%% depreciated in this version as it is no longer necessary. AASTeX 
%% automatically takes care of all commas and "and"s between authors names.

%% AASTeX 6.2 has the new \collaboration and \nocollaboration commands to
%% provide the collaboration status of a group of authors. These commands 
%% can be used either before or after the list of corresponding authors. The
%% argument for \collaboration is the collaboration identifier. Authors are
%% encouraged to surround collaboration identifiers with ()s. The 
%% \nocollaboration command takes no argument and exists to indicate that
%% the nearby authors are not part of surrounding collaborations.

%% Mark off the abstract in the ``abstract'' environment. 
\begin{abstract}

%{\color{green} Manuscript section (Maximum)\\
%Abstract length:	250 words\\
%References:	50\\
%Total Figures and Tables:	5\\
%Manuscript length: 3500 words\\
%Machine readable tables: 1}\\\\

% We have identified 127 HMXBs in the Small Magellanic Cloud (SMC), a nearby star-forming galaxy, as part of a Chandra X-ray Visionary Program (XVP). We use these, {\bf \textcolor{red}{in addition to 14 previously-known HMXBs in the SMC without \chandra detections}} to investigate the formation efficiency of young accreting binaries in the low metallicity environment of the SMC. 

We have compiled the most complete census of High-Mass X-ray Binaries (HMXBs) in the Small Magellanic Cloud with the aim to investigate the formation efficiency of young accreting binaries in its low metallicity environment. In total, we use 127 X-ray sources with detections in our \chandra X-ray Visionary Program (XVP), supplemented by 14 additional (likely and confirmed) HMXBs identified by \cite{2016A&A...586A..81H} that fall within the XVP area, but are not either detected in our survey (9 sources) or matched with any XVP source that has at least one OB counterpart in the OGLE-III catalog  (5 sources). Specifically, we examine the number ratio of the HMXBs [N(HMXBs)] to {\it (a)} the number of OB stars, {\it (b)} the local star-formation rate (SFR), and {\it (c)} the stellar mass produced during the specific star-formation burst, all as a function of the age of their parent stellar populations. Each of these indicators serves a different role, but in all cases we find that the HMXB formation efficiency increases as a function of time (following a burst of star formation) up to $\sim$40--60\,Myr, and then gradually decreases. The peak formation efficiency N(HMXB)/SFR is (49 $\pm$ 14) $[10^{-5}~{\rm M_{\odot}/yr}]^{-1}$, in good agreement with previous estimates of the average formation efficiency in the broad $\sim$20--60\,Myr age range. The frequency of HMXBs is a factor of 8$\times$ higher than at $\sim$10\,Myr, and 4$\times$ higher than at $\sim$260\,Myr, i.e. at earlier and later epochs, respectively.

\end{abstract}

%% Keywords should appear after the \end{abstract} command. 
%% See the online documentation for the full list of available subject
%% keywords and the rules for their use.

%% From the front matter, we move on to the body of the paper.
%% Sections are demarcated by \section and \subsection, respectively.
%% Observe the use of the LaTeX \label
%% command after the \subsection to give a symbolic KEY to the
%% subsection for cross-referencing in a \ref command.
%% You can use LaTeX's \ref and \label commands to keep track of
%% cross-references to sections, equations, tables, and figures.
%% That way, if you change the order of any elements, LaTeX will
%% automatically renumber them.
%%
%% We recommend that authors also use the natbib \citep
%% and \citet commands to identify citations.  The citations are
%% tied to the reference list via symbolic KEYs. The KEY corresponds
%% to the KEY in the \bibitem in the reference list below. 

\section{Introduction} \label{sec:intro}

X-ray binaries (XRBs) are our main tool for studying the populations of compact objects in galaxies, and the formation and evolution of intermediate and higher mass binary stellar systems. Systematic studies of nearby galaxies have provided initial estimates of the formation rate of XRBs as a function of the SFR and stellar mass (M$\star$) of their host galaxies (\citealt{2010ApJ...724..559L}, \citealt{2012MNRAS.419.2095M}). A strong dependence of the XRB formation rate on  age and metallicity has been predicted (\citealt{2006MNRAS.370.2079D}, \citealt{2010ApJ...725.1984L}, \citealt{2013ApJ...764...41F}); however, we are only now starting to probe the details of this connection (\citealt{2005A&A...431..597S}, \citealt{2010ApJ...716L.140A}, \textcolor{green}{\citealt{2013ApJ...772...12W}}, \citealt{2016MNRAS.459..528A}). 
The Small Magellanic Cloud (SMC) is the ideal target to study the dependence of the formation efficiency of XRBs on the age of their parent stellar population. It is our second nearest (D=61.9$\pm$0.6 kpc;  \citealt{2015AJ....149..179D}) star-forming galaxy, and offers a clear picture of its spatially-resolved star-formation (SF) history (\citealt{2004AJ....127.1531H}, hereafter [HZ04]). Furthermore, it has low metallicity (\citealt{1998AJ....115..605L}; \citealt{2016MNRAS.459..528A}, and references therein), and hosts one of the largest known populations of High-Mass X-ray Binaries (HMXBs) known in a galaxy, rivaling that of the Milky Way (e.g. \citealt{2015MNRAS.452..969C}; \citealt{2016A&A...586A..81H}, hereafter  [HS16]).

In order to study the HMXB populations in the SMC in detail, and in particular their connection with their parent stellar population, we performed a deep \chandra {\it X-ray Visionary Project  (XVP)} survey of selected SMC regions chosen to sample stellar populations of different ages. The first results on the spectral and timing properties of pulsars detected in the survey fields are presented in \cite{2016ApJ...826....4H,2017ApJ...847...26H}.  In this work, we present our measurement of the formation efficiency of the SMC HMXB populations as a function of their age, the most detailed such measurement so far.  In \S \ref{sec:survey} we describe briefly the \chandra  XVP SMC survey, and the source sample used, while in \S \ref{sec:SFHs} we discuss the SF history of the regions studied, and we link the SF in each region with the XRB populations. In \S \ref{sec:FormEff} we estimate the formation efficiency of  HMXBs, and present the HMXB delay time distribution (DTD). In \S \ref{sec:Disc} we discuss these results and compare the different HMXB formation efficiency indicators. The most important findings are summarized in \S \ref{sec:Conclusions}.

Throughout this work, we adopt a distance modulus of $(m - M)_{V} = 18.96\pm0.02$ mag \citep{2015AJ....149..179D}, $R_{V} =  2.74 \pm0.13$ \citep{2003ApJ...594..279G}, and $E(B - V) (= A_{V}/R_{V}) = 0.09\pm0.02$ mag \citep{1999AcA....49..437U}, thus the extinction $A_{V}$ is estimated to be 0.25~mag, and in turn\footnote{Assuming the extinction curve of \cite{1989ApJ...345..245C}.} $A_{I}=0.12$ mag, and $E(V - I) = 0.13$ mag.

\section{Survey description and data analysis} \label{sec:survey}

The SMC has been surveyed extensively in the X-ray band with \einstein (e.g. \citealt{1981ApJ...243..736S}), \rosat (e.g. \citealt{2000A&AS..147...75S}, \citealt{2000A&AS..142...41H}), \rxte (e.g. \citealt{2005ApJS..161...96L}, \citealt{2009IAUS..256..361C}), \asca (e.g. \citealt{2003PASJ...55..161Y}), and \xmm (e.g. \citealt{2012A&A...545A.128H}, \citealt{2013A&A...558A...3S}), with the latter yielding the most extensive survey of the galaxy down to luminosities of $5 \times 10^{33}$\ergs\, \citep{2012A&A...545A.128H}. In order to reach well within the regime of the X-ray emission of quiescent XRBs ($\sim10^{32}$\ergs) and obtain as complete a picture of its HMXB populations as possible, a \chandra XVP Program was awarded (PI A.~Zezas) to perform a comprehensive survey, to a depth of 100~ks exposure, of 11 fields selected to represent young ($<$ 100\,Myr) stellar populations of different ages. These observations were performed from  December 2012 to February 2014, utilizing the ACIS-I imaging mode. 

\begin{figure}[htbp] %  figure placement: here, top, bottom, or page
   \centering
   \includegraphics[width=2.65in,angle=270]{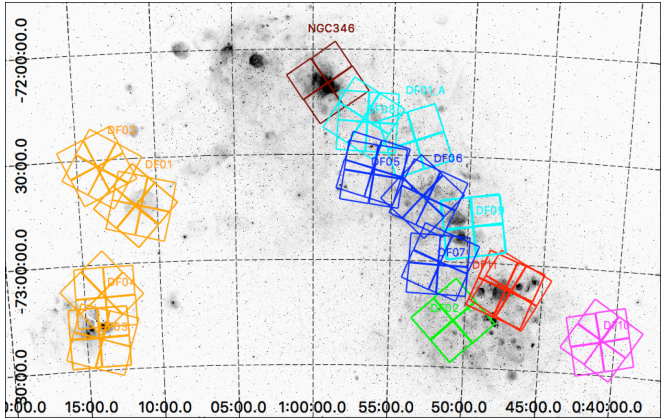}
\vspace{-0.5cm}
   \caption{MCELS \Ha image (F. Winkler/Middlebury College, the MCELS Team, and NOAO/AURA/NSF) overlaid with the 14 \chandra fields analyzed in this work, color-coded for the ages of their parent stellar population  
 (orange: 11\,Myr, blue: 34\,Myr, cyan: 42\,Myr, magenta: 67\,Myr). Three fields have two distinct stellar populations: DF11 (7\,Myr and 42\,Myr -- red), DF02\_A (42\,Myr and 167.9\,Myr -- green), and NGC346 (5\,Myr and 42\,Myr -- maroon).}\label{fig:fields}
\end{figure}

In addition, we also analyzed 3 archival observations reaching the same 100~ks depth. Two of these fields (PI A. Zezas; observed in 2006) overlap partially with fields from the XVP survey, and the third is centered on NGC~346 (PI M. Corcoran; observed in 2001). Although analyses of these data have been presented elsewhere (\citealt{2010ApJ...716.1217L} and \citealt{2002ApJ...580..225N}, respectively), we opted to reanalyze all consistently using the latest calibrations. In Figure \ref{fig:fields} we present a Magellanic Cloud Emission Line Survey (MCELS) \Ha image of the SMC overlaid with the observed fields, color-coded for the age of their stellar population derived using data from [HZ04]. 

Source detection employed {\em CIAO} {\sc wavdetect} in 4 bands (broad 0.5--7.0 keV, soft 0.5--1.2 keV, medium 1.2--2.0 keV, and hard 2.0--7.0 keV) on all individual ObsIDs and the merged dataset for each field. All subsequent data analysis steps (source photometry, screening, spectral fitting, timing analysis)  were performed with {\em ACIS Extract} \citep[AE Version 2014may23;][]{Broos...et...al...2010,2012ascl.soft03001B}. This yielded 2,393 sources down to a limiting flux of $2.6 \times 10^{-16}$ \funit\, in the full (0.5--8.0 keV) band ($\sim$50\% complete at $7.94 \times 10^{-16}$ \funit). Further details on the survey, data analysis, and the complete source catalog\footnote{The limit we  set on the probability of these sources of being just a background fluctuation -- value of PROB\_NO\_SOURCE or ${\rm P_{B}}$ in {\em AE} -- is 0.01.} are presented in Antoniou et al. (2019, in prep).

In order to identify the HMXBs in the complete source catalog, we cross-correlated the locations of the X-ray sources with the OGLE-III catalog of stars in the SMC \citep{2008AcA....58..329U}. We used a cross-correlation radius based on Eq.\,5 of  \cite{2005ApJ...635..907H}, limited to a minimum radius of 1\arcsec\, based on the minimum combined positional uncertainty of the X-ray and optical catalogs.
 Following \citet{2010ApJ...716L.140A} and \citet{2016MNRAS.459..528A}, we classified as HMXBs X-ray sources with optical counterparts within the OB-star locus of the ($V, V-I$) color-magnitude diagram (CMD). This locus is based on the location of massive stars from the spectroscopic census of the SMC  \citep{2010AJ....140..416B}.  In order to account for the well-known effect that Be-XRBs (the most numerous subclass of SMC HMXBs) appear redder than OB stars due to the circumstellar disk of their Oe/Be star  (e.g. \citealt{2009ApJ...697.1695A}, \citealt{2009ApJ...707.1080A}, \citealt{2010AJ....140..416B}),  we extended the locus to redder colors based on the locations of known HMXBs from [HS16]\footnote{Using the  120 high-confidence sources reported in [HS16].} on the same CMD.  We define the "extended" OB-star locus (hereafter referred simply as the OB-star locus) to lie within $V\leq18$ mag, $I\leq18$ mag,  and $-0.4\leq V-I\leq0.6$ mag (Figure~\ref{fig:DF11cparts}).

\section{Star-formation history} \label{sec:SFHs}

A spatially-resolved SF history was derived for the SMC with a scale of $12\arcmin \times 12\arcmin$ (216.2~pc $\times$ 216.2~pc) as part of the Magellanic Clouds Photometric Survey (MCPS [HZ04]). The SF history in each of the \chandra fields we adopt is the total SF history of the MCPS subregions they encompass, weighted by the fraction of the area of each MCPS subregion covered by the field.

\begin{figure}
%\begin{figure}[t] 
\centering
%\hspace{-1.5cm}
\includegraphics[width=0.45\textwidth]{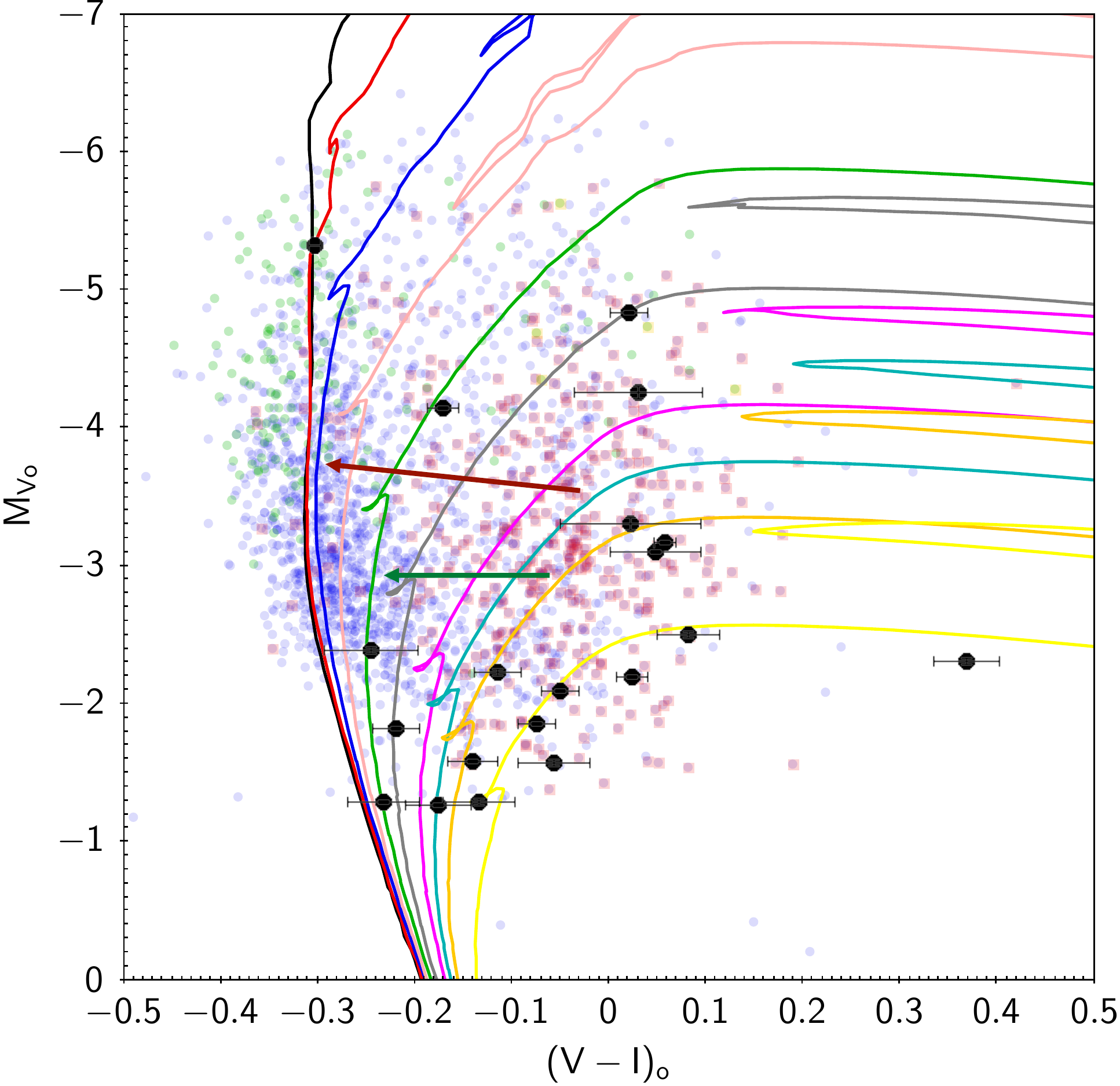}
\caption{An example of the age determination for the HMXBs identified in field DF11 of the \chandra XVP survey. The extinction-corrected magnitude and color are defined as  $M_{V_o} = (m-M)_{V} - A_{V} = V - 18.96 - 0.25$, and $(V-I)_{o} = (V-I) - E(V-I) = (V-I) - 0.13$, respectively. The underlying points are from the spectroscopic census of OB stars in the SMC (blue: B stars; green: O stars, red: Be stars, orange: Oe stars) by \cite{2010AJ....140..416B}, overlaid with the PARSEC isochrones (black: 4.5\,Myr; red: 6.3\,Myr; blue: 10.0\,Myr; pink: 15.8\,Myr; green: 25.1\,Myr; grey: 39.8\,Myr; magenta: 63.1\,Myr; cyan: 79.4\,Myr; orange: 100.0\,Myr; yellow: 158.5\,Myr). The two vectors, maroon and dark green, show the shift of the Be stars due to their intrinsic reddening with respect to the B stars in the [B0,B2) and [B2,B4) spectral-type bins, respectively (the start and end points of the arrows are centered at the median $M_{V_o}$ and $(V-I)_{o}$ values of each population).}\label{fig:DF11cparts}
\end{figure}

In order to estimate the chance coincidence probability of identifying spurious matches from the OGLE-III catalog as the optical counterparts of the HMXBs, we performed extensive Monte Carlo simulations  following \cite{2009ApJ...697.1695A}, and \cite{2016MNRAS.459..528A}.  These results indicate that for a search radius of 1\arcsec, about 10\% of the bright blue ($V\leq16$ mag and $-0.4\leq(V-I)\leq0.6$ mag) matches are spurious associations, with this probability increasing to 17\% and 79\% for $16<V\leq17$ mag, and $17<V\leq18$ mag, respectively (Antoniou et al. 2019, in prep.). These results show  that the brightest (and subsequently bluest, for objects of similar brightness) match for sources with multiple matches is the most likely optical counterpart (c.f. \citealt{2009ApJ...697.1695A}). Our final HMXB sample consists of candidate sources identified in this work (127 sources matched with a total of 143 early-type stars), supplemented by all additional HMXBs identified by [HS16] (likely and confirmed sources) that fall within the XVP area, but are not either detected in our survey (9 sources) or matched with any XVP source that has at least one OB counterpart in the OGLE-III catalog  (5 sources)\footnote{Out of the  120 high-confidence sources reported in [HS16], only 65 fall within the 14 \chandra fields used in this work. Also, only 41 out of the 127 HMXB candidates in our survey have a match in the catalog of [HS16], i.e. 86 candidate HMXBs from this work are not listed in [HS16]. These sources have a limiting ${\rm L_{X}}$ of $\sim 1.7\times10^{32}$ erg/s. Five  additional sources from [HS16] are not included in our final list because they have only \rxte or \integral positions with $\gtrsim$100\arcsec\, uncertainties.}. Our final source sample comprises 141 candidate and confirmed HMXBs.

For the purpose of measuring the formation rate of HMXBs with respect to their parent stellar populations, when an HMXB falls within two or more overlapping \chandra fields, we associate it with the field that has a peak of SF at a look-back time consistent with its age (indicated in Column 4 of Table~\ref{tbl:ageHMXBs}). This is necessary because we are measuring the SF history of each \chandra field.

\begin{table*}[h]
 \centering
%\begin{minipage}{85mm}
\caption{HMXB Populations and Formation Efficiency}\label{tbl:ageHMXBs}
 \begin{tabular}{@{}rlrrrrrrrr@{}}
  \hline
    \hline
  Age &   Field    & HMXBs & \multicolumn{4}{c}{SF burst} &  \multicolumn{3}{c}{HMXB Formation Efficiency}  \\
     \cline{4-7} \cline{8-10}
  Bin &     ID       &            &  Age     &  Span & Rate    &   M$\star$ & SFR    &  OB stars   & M$\star$ \\
   ${\rm [Myr]}$  &    &    &  ${\rm [Myr]}$   &  ${\rm [Myr]}$  &  [10$^{-3}$~\msunsimple/yr]  &  [10$^{5}$~\msunsimple]  &  $[10^{-1}~{\rm M_{\odot}/yr}]^{-1}$  &  [$10^{-4}$]  & $[10^{-6}~{\rm M_{\odot}}]^{-1}$\\
\colnumbers 
\hline
6  & DF11      & 1     &   7    & 4             &  $115_{-53}^{+46}$   &  $3.6_{-1.7}^{+3.0}$ &  $1.3_{-1.3}^{+1.6}$  &  $4.3\pm4.3$   &  $8.5_{-8.5}^{+10.6}$ \\
     & NGC346    & 1     &   5    & $\sim$1--2      &  $56_{-55}^{+40}$    &  $0.7_{-0.7}^{+0.5}$ &  &  &   \\ 
\hline
11 & DF01      & 1     &  11    & 6             &  $37_{-10}^{+11}$    &  $2.1_{-0.6}^{+1.0}$ & $5.5\pm3.2$ &  $63\pm33$   &  $7.5_{-4.4}^{+4.9}$\\  
     & DF02      & 0     &  11      & 6             &  $80\pm14$    &  $4.5_{-1.2}^{+2.1}$ &  &  &      \\
     & DF03      & 5     &  11      & 6             &  $73_{-12}^{+10}$    &  $3.9_{-0.8}^{+1.0}$ &  &  &    \\
     & DF04      & 4     &  11      & 6             &  $57\pm8$      &  $3.2_{-0.6}^{+1.1}$ &  &  &      \\  
\hline
34 & DF05      & 5     &  34 & 43 & $30_{-5}^{+6}$ & $15_{-3}^{+4}$ & $24_{-11}^{+10}$ &  $34\pm12$   &  $5.4_{-2.6}^{+2.7}$\\  
     & DF06      & 11    & 34  & 36         &  $37_{-7}^{+9}$      &  $16_{-4}^{+6}$ &  &  &    \\
     & DF07      & 11    & 34  & 34 & $42_{-17}^{+10}$ & $18_{-8}^{+7}$ &  &  &     \\
\hline
42 & DF08      & 24    & 42     & 43           &  $38\pm5$              &  $18_{-2}^{+3}$ & $49\pm14$ &  $58\pm15$   &  $12_{-4}^{+5}$\\
     & DF09      & 7     &   42       & 28                         &  $33\pm6$      &  $9.4_{-2.1}^{+3.4}$  &  &  &  \\
     & DF11      & 16    &   42       & 22           &  $53_{-10}^{+9}$     &  $16_{-5}^{+7}$  &  &  &  \\
     & DF01\_A   & 18    &   42       & 30           &  $37\pm3$              &  $13\pm2$  &  &  &  \\
     & DF02\_A   & 27    &   42       & 41           &  $25\pm5$      &  $13_{-5}^{+6}$  &  &  &    \\
     & NGC346    & 7     &   42       & 30           &  $37_{-4}^{+5}$      &  $11_{-2}^{+3}$  &  &  &  \\
\hline
67 & DF10      & 2     & 67 & 29 &  $2.7_{-2.5}^{+6.9}$       &  $7.2_{-2.2}^{+3.4}$ & $74_{-74}^{+197}$ &  $25\pm18$   &  $2.8_{-2.1}^{+2.4}$\\
\hline
266 & DF02\_A  & 1 & 266 & 436 & $8.3_{-1.7}^{+1.9}$ & $68_{-7}^{+9}$ & $12\pm12$ &  $3.2\pm3.2$   &  0.1$\pm$0.1\\
\hline 
\hline 
\end{tabular}
\tablecomments{
Column (1): Average age (using values in Column 4) for the stellar populations in a given SF episode; Column (2): Field ID (Figure~\ref{fig:fields}); Column (3): Number of HMXBs in each field associated with the respective SF episode; Columns (4) and (5): age and time-span (FWHM) of the dominant SF episode; Column (6): peak SFR of this episode (errors are based on the upper and lower SFR ranges reported by  [HZ04]); Column (7):  Total stellar mass (M$\star$) produced in the SF episode (based on the integration of the SFR time-evolution); Columns (8), (9), (10): The HMXB formation efficiency based on the ratio of N(HMXBs) (Column 3) to the SFR (Column 6), the N(OBs), and the stellar mass (Column 7) produced during the SF burst they are associated with (see \S\ref{sec:SFHs}).
}
%\end{minipage}
\end{table*}

\begin{figure} 
 %  figure placement: here, top, bottom, or page
   \centering
\includegraphics[width=3.25in,angle=0]{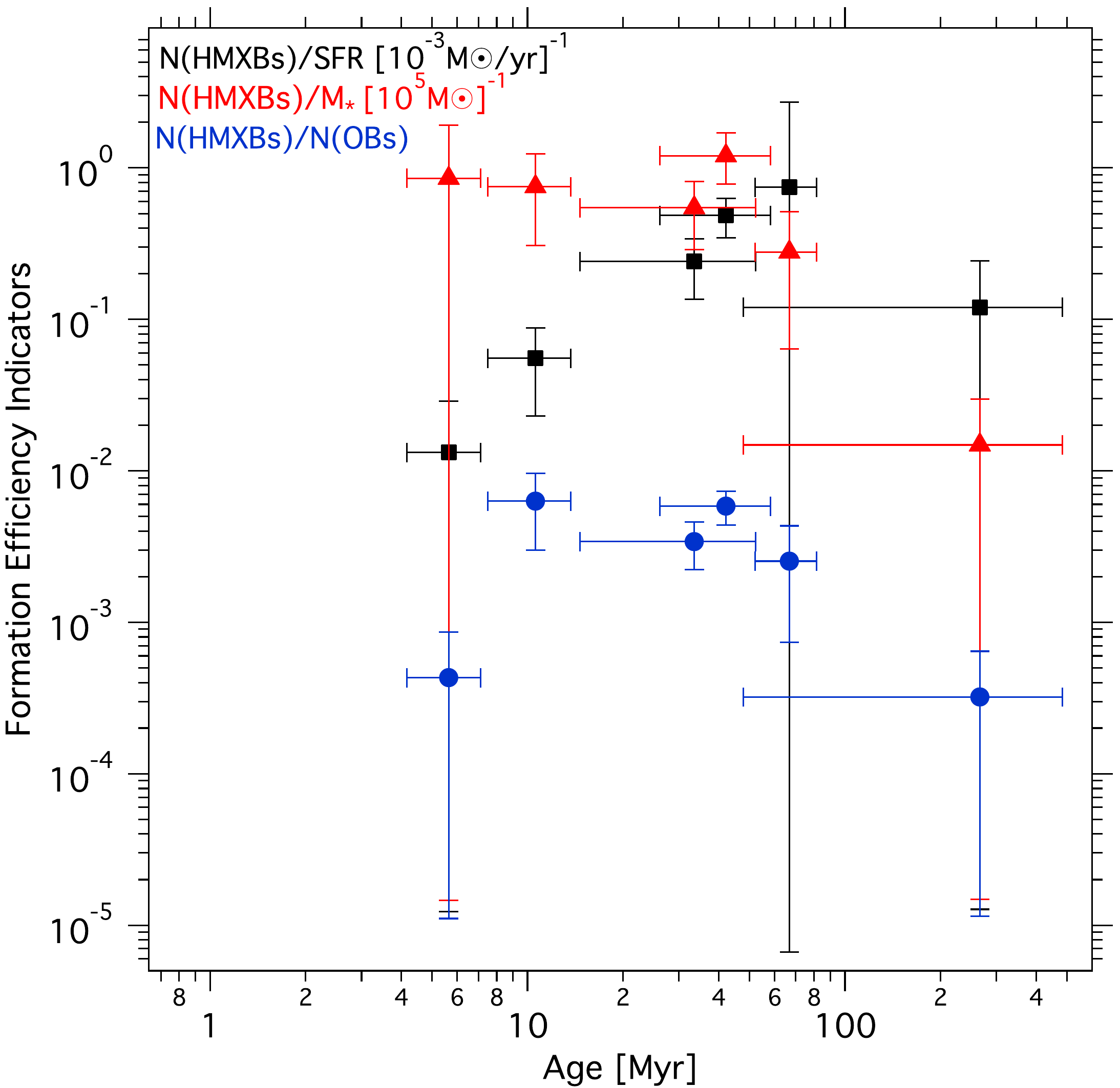}
\caption{The following formation efficiency indicators  are shown as a  function of the age of their parent stellar population: N(HMXBs) over the SFR (black squares); N(HMXBs) over the stellar mass produced during the major SF burst (red triangles); N(HMXBs)/N(OBs) in the studied fields (blue circles).
}\label{fig:allFE}
\end{figure}

Since our goal is to measure the formation rate of HMXBs as a function of the age of their parent stellar population, we first need to constrain the ages of the HMXBs and associate them with individual SF episodes responsible for the birth of their progenitors. Ages are derived from optical counterpart positions on the ($V, V-I$) CMD with respect to the PARSEC isochrones (v1.2S; \citealt{2012MNRAS.427..127B}) generated by CMD 2.8\footnote{http://stev.oapd.inaf.it/cmd} for ${\rm Z=0.004}$ (Figure~\ref{fig:DF11cparts}). We note that due to their circumstellar disks the Be stars are intrinsically redder than B-type stars of the same spectral type. By comparing the $(V-I)$ colors of B and Be stars from the census of \cite{2010AJ....140..416B}, we have found that early [B0--B2), and mid [B2--B4) spectral-type Be stars (i.e. within the typical range of companions of known SMC HMXBs; e.g \citealt{2008MNRAS.388.1198M}, \citealt{2009ApJ...707.1080A}, \citealt{2014MNRAS.438.2005M}) have reddening-corrected $(V-I)_{o}$ colors $\sim$0.3 and $\sim$0.2\,mag redder, respectively, than B stars of the same spectral type\footnote{For spectral types later than B4, the very small size of the Be star sample does not allow us to derive any conclusions.}. This systematic displacement (also obvious in Figure~\ref{fig:DF11cparts}) implies that a Be system appears redder than a B star due to its equatorial disk.

Furthermore, each optical counterpart is associated with a SF episode taking into account the fact that stars can be associated with a SF event that overlaps with the age-range of isochrones consistent with its location on the CMD. For example, the SF history of \chandra field DF11 shows two prominent peaks at 7\,Myr and 42\,Myr. Out of the 17 HMXBs of DF11 (14 with unique optical counterparts and 3 with two optical matches; Column 3 of Table~\ref{tbl:ageHMXBs}) only one X-ray source has an optical counterpart with a location on the  OGLE-III ($V, V-I$) CMD consistent  with the peak at 7\,Myr, while the remaining 16 have ages consistent the peak at 42\,Myr (Figure~\ref{fig:DF11cparts}).
 The HMXB and SF age associations, together with the SFR, duration and total stellar mass of the relevant SF episode, are presented in Table~\ref{tbl:ageHMXBs}.  The stellar mass formed during the SF episode associated with each HMXB population is calculated by integrating the SF history during the period of the star-formation episode of interest in each relevant field. For simplicity we approximate each SF episode as a sequence of Gaussian events; generally 1-3 Gaussians are adequate to reproduce the evolution of the SFR during a SF event.

\section{HMXB formation efficiency} \label{sec:FormEff}

We derive three different metrics of the age-dependent formation efficiency of HMXBs, the number of HMXBs in different ages with respect to the: {\it (a)} number of OB stars, N(OBs), in their respective \chandra field; {\it (b)} SFR of their parent stellar population; and {\it (c)} stellar mass formed during the SF episode they are associated with.
The OB stars are from the OGLE-III catalog (\S\ref{sec:SFHs}), while the SFRs as a function of age are from [HZ04] and are listed in Column 6 of Table~\ref{tbl:ageHMXBs}. The age-dependence of these three different tracers of the HMXB formation rate is shown in  Figure~\ref{fig:allFE}. We have grouped together \chandra fields that have similar ages (as indicated by the different group of fields shown in Table~\ref{tbl:ageHMXBs}; e.g. DF05, DF06, and DF07, all showing a prominent peak in their SF histories at $\sim$34 Myr). The error bars in the x-axis indicate the average age range of the stellar populations in each age bin.

The N(HMXBs)/N(OBs) ratio  shows a steep increase (by about an order of magnitude) from $\sim$6 to 10\,Myr, then a flattening up to $\sim$60\,Myr, followed by a drop (by about an order of magnitude again) for ages older than $\sim$60\,Myr.

On the other hand, the N(HMXBs)/SFR ratio increases rapidly up to $\sim$40--60\,Myr, and then gradually decreases for older stellar populations. This result is consistent with previous lower age-resolution small-scale studies (involving shallow \chandra and \xmm observations), which show an increased formation efficiency of HMXBs at ages between 30--60\,Myr (of fields across the SMC Bar) compared to younger stellar populations (SMC Wing) (\citealt{2005A&A...431..597S}; \citealt{2009ApJ...697.1695A}).

The N(HMXBs)/M$\star$ ratio instead remains flat up to $\sim$60\,Myr, but it decreases for later ages as well. The data points averaged in the second age bin (at 11~Myr) correspond to fields that lie across the SMC Wing (Figure~\ref{fig:fields}), one of the youngest identified SMC regions [HZ04]. These fields, although they have large SFRs (Table~\ref{tbl:ageHMXBs}), they show SF episodes of small duration. This results in a small stellar mass formed, resulting in turn in a large N(HMXBs)/M$\star$ ratio.  The same argument holds for the similar behavior of the N(HMXBs)/N(OBs) ratio (which is effectively a proxy for stellar mass in massive stars).

An alternative formulation of the time taken for the formation of a class of astronomical objects (in this case, HMXBs) from the SF event that gives rise to its progenitor stellar systems is described by the Delay Time Distribution (DTD). The DTD is defined as the production rate of objects as a function of time after an hypothetical brief SF burst. \cite{2015ApJ...804L..25B} (hereafter [B15])  described a method to recover the DTD from an object catalog and a SF history map, and applied it to LMC planetary nebulae. Here, we apply the same method to the SMC HMXB catalog described in \S \ref{sec:SFHs}. The only difference with respect to the [B15] analysis is that the \textit{Chandra} fields we used to derive the HMXB catalog do not cover a contiguous or uniform part of the SMC, and in many cases there is only partial overlap between a given \textit{Chandra} field and a specific MCPS subregion. For this reason, we multiplied the SFR of each MCPS subregion in the SF history map of [HZ04] by a weight between 0 and 1, which  represents the fraction of the surface area of the subregion covered by \textit{Chandra} (as was done in \S \ref{sec:SFHs}).

\begin{figure}[htbp] %  figure placement: here, top, bottom, or page
   \centering
   \includegraphics[width=3.25in]{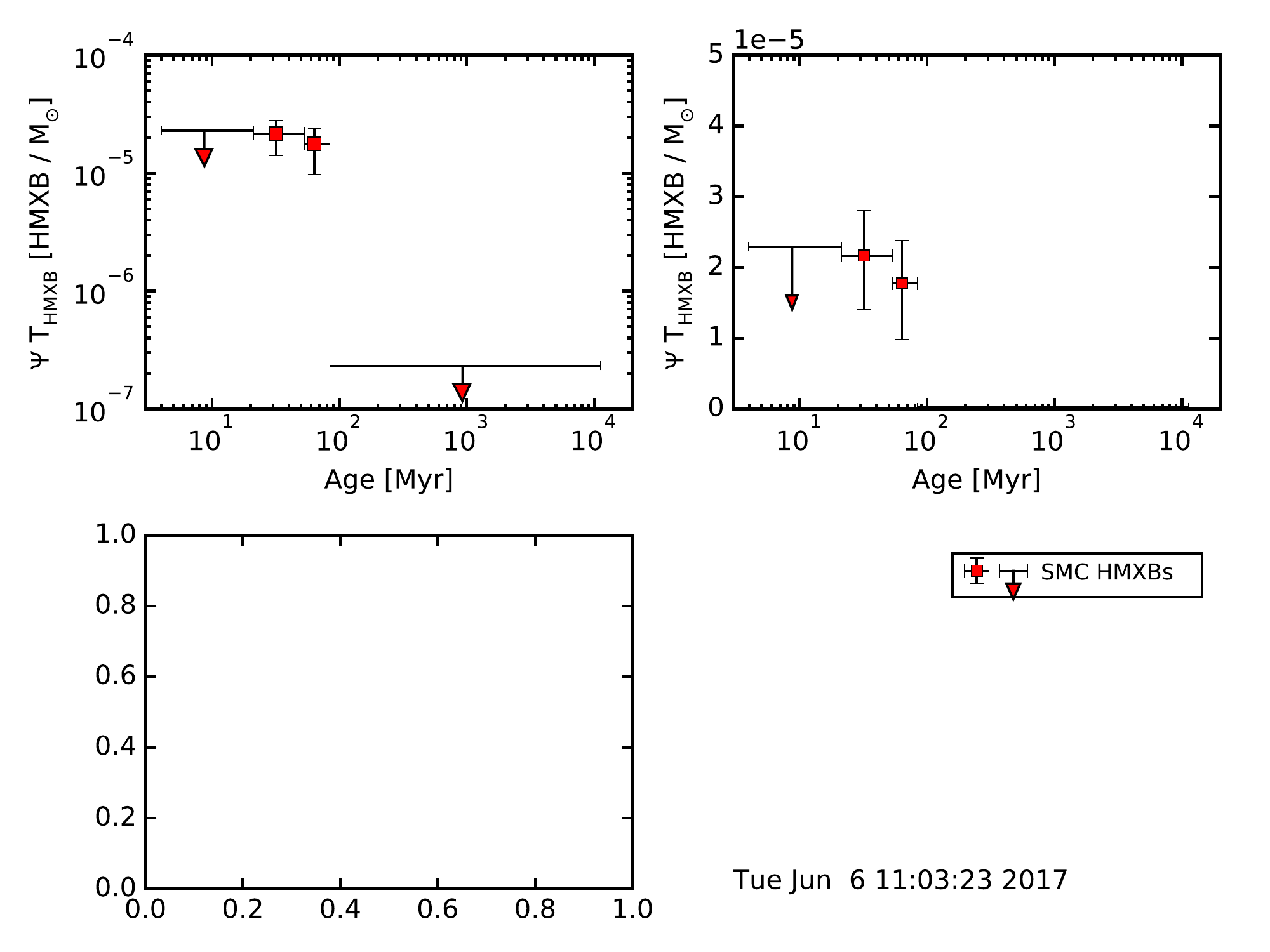} 
   \caption{HMXBs Delay Time Distribution (following [B15]).}\label{fig:DTD}
\end{figure}

The resulting DTD is presented in Figure~\ref{fig:DTD}. We have used the temporal binning that offers the best compromise between DTD resolution and detection significance, given N(HMXBs) and the native resolution of the SF history map. We detect significant signal in the DTD of HMXB progenitors for stellar ages 21--53\,Myr, and 53--84\,Myr. Stellar populations in this age range generate $\sim2\times10^{-5}$ HMXBs per unit stellar mass. This formation efficiency, $\Psi {\rm T_{HMXB}}$, is the product of the specific HMXB formation rate, $\Psi$ [HMXBs yr$^{-1}$ \msunsimple$^{-1}$], and 
mean HMXB lifetime, ${\rm T_{HMXB}}$ [yr] ---see Eqs.~(1)--(3) and \S 2 in [B15]. For stellar populations younger than 21\,Myr, we obtain a shallow 2$\sigma$ upper limit to the HMXB formation efficiency of $\sim2.3\times10^{-5}\,$\msunsimple$^{-1}$. For stellar populations older than 84\,Myr, we obtain a much lower upper limit of $2.5\times10^{-7}\,$\msun$^{-1}$. This indicates that there must be a maximum delay time for HMXB formation of less than 84\,Myr, but longer than 53\,Myr, given the significant detection in this bin.

\section{Discussion}\label{sec:Disc}

In section \S \ref{sec:FormEff} we presented an analysis of the formation efficiency of HMXBs in the SMC based on a set of deep \chandra observations of this galaxy. We calculate this  based on three different indicators, (N(HMXBs)/SFR, N(HMXBs)/N(OBs), and N(HMXBs)/M$\star$, all as a function of the age of the major associated SF burst), as well at the delay function formulation. We find that there is an increase in the formation rate for ages $\gtrsim$10--20\,Myr and up to 40--60\,Myr followed by a decline at older ages. The three HMXB formation efficiency indicators presented in Figure~\ref{fig:allFE} serve different purposes. 

N(HMXBs)/N(OBs) is observationally driven, and takes into account the present-day numbers of OB stars. However, it does not take into account the donor star rejuvenation due to the first mass transfer from the initially more massive star that subsequently explodes as a SN and leaves behind a neutron star (or a black hole). Because of this rejuvenation, the system will live longer than single stars of similar mass formed during the same SF episode. This discrepancy is smaller for SF episodes of similar or longer duration compared to the lifetime of HMXB systems. Nonetheless, the N(HMXBs)/N(OBs) ratio is an indicator that can be calculated directly for any nearby galaxy with resolved stellar populations, without the need to derive their SF history. Therefore, it serves as a useful proxy of the relative formation rate of HMXBs that can be applied to large samples of galaxies.

N(HMXBs)/SFR is based on the SF episode of the parent stellar population. It takes into account the SF event that created the binaries we observe today, but not the duration of the SF burst. In this work, we derived a peak formation efficiency N(HMXB)/SFR of (49 $\pm$ 14) $[10^{-5}~{\rm M_{\odot}/yr}]^{-1}$, in good agreement with previous estimates of the average formation efficiency in the broad $\sim$20--60\,Myr age range. We also find a factor of 8 rise in the formation efficiency with respect to younger populations ($\sim$10\,Myr) and a factor of 4 decline in older epochs ($\sim$260\,Myr). The different behavior of the N(HMXBs)/SFR with respect to the N(HMXBs)/N(OBs) indicator could be the result of the age dependence of N(OBs): as the stellar populations age, a smaller number of OB stars is expected to be present. From a simple stellar lifetime argument folded through the IMF,  the number of OB stars will be reduced with time, while the rejuvenation of the donor star would result in a longer lifetime of the binary systems. While N(HMXBs)/SFR is considered a more accurate representation of the formation efficiency of young accreting binaries than  N(HMXBs)/N(OBs), it is similarly problematic for providing observational constraints in sophisticated population synthesis models (e.g. \citealt{2018ApJS..237....1A}).

More suitable is the ratio of N(HMXBs) to the total stellar mass produced in the relevant SF burst, since this takes into account the SF burst duration (the integral of the SFR as a function of time). This is the fundamental relation that we were aiming to derive from this \chandra XVP program, and the one that best resembles the delay function of the HMXBs. In fact, our N(HMXBs)/M$\star$ indicator (black points in Figure~\ref{fig:allFE}) is consistent (within the errors) with the DTD (shown in Figure~\ref{fig:DTD}). Moreover, our results are in good qualitative agreement with the simulated stellar-mass normalized total X-ray luminosity output of a galaxy as a function of age of \cite{2013ApJ...764...41F}, who find an increase at ages $\gtrsim20$\,Myr and a decrease at ages $\lesssim80$\,Myr. This effect becomes more prominent in metallicities like those of the SMC (1/5\zsun).

The \chandra fields along the SMC Wing (DF01--DF04 in Figure~\ref{fig:fields}) produce only a small number of HMXBs based on the surveys conducted so far (\citealt{2008MNRAS.383..330M}; Antoniou et al., in prep.). We attribute this deficiency on the strong but very recent ($<$10\,Myr) star formation of the fields in this area \citep{2010ApJ...716L.140A} compared to the SMC Bar regions (typically $\sim$25-60\,Myr). Although this deficit might indicate an elusive, young, population of HMXBs, such as highly absorbed HMXBs (e.g. \citealt{2015A&ARv..23....2W}), based on XRB evolution models we would expect a small number of XRBs at such young ages as only the most massive systems would have produced compact objects \citep{2008ApJS..174..223B}. Because of the large mass of the progenitors of these systems and the low metallicity of the SMC, we would expect these systems to be predominantly black-hole XRBs \citep{2010ApJ...716L.140A}.

The time-resolved HMXB formation efficiency with respect to the stellar mass presented in Figure~\ref{fig:allFE} is in good agreement with the general trend estimated by \cite{2007AstL...33..437S}, who find a peak at  similar ages ($\sim$40\, Myr; albeit with coarser time resolution).  We attribute differences in the absolute value of the formation efficiency in the two works to the fact that \cite{2007AstL...33..437S} consider only massive stars (M $>$ 8\msunsimple) in their calculation of stellar mass that was used to normalize the number of HMXBs.

Our results are also in agreement with studies of the formation efficiency of massive Oe/Be stars in the Magellanic Clouds (e.g. \citealt{2006A&A...452..273M}, \citealt{2007A&A...462..683M}, \citealt{2009AJ....138.1003B}, \citealt{2010AJ....140..416B}), and the Milky Way \citep{2005ApJS..161..118M}. 
These show a peak at ages of $\sim$20--50\,Myr, matching the age of maximum production of HMXBs at least at the metallicity of the SMC. This similarity could indicate  that: {\it (a)}  the Be stars, the donor stars of Be-XRBs (the predominant HMXB population in the SMC), are the result of binary evolution (e.g. \citealt{2003PASP..115.1153P}, and references therein), and/or {\it (b)} the larger mass-loss rates of Be stars through their equatorial winds (in comparison to the much weaker spherical stellar winds) lead to an enhanced population of active XRBs (c.f. \citealt{2010ApJ...716L.140A}). However, only detailed population synthesis models accounting for the complex orbital evolution and mass-transfer in eccentric binaries (e.g. \citealt{2006MNRAS.370.2079D}) can distinguish between these possibilities.

Finally, a first assessment of the overall XRB formation rate in the LMC, which has two SF episodes at similar ages as the SMC ($12.6$\,Myr and $42$\,Myr) but with different intensities, indicates that the formation efficiency of its overall XRB population is $\sim$17 times lower than in the SMC \citep{2016MNRAS.459..528A}. This could be the result of a metallicity effect (e.g. Be stars form more efficiently at lower metallicities as shown by \citealt{2007A&A...472..577M} and \citealt{2013MNRAS.435.3103I}). Furthermore,  \cite{2006MNRAS.370.2079D} finds that at the $\sim1/5$\zsun\, metallicity of the SMC, population synthesis models predict 3 times larger populations of HMXBs than in the Milky Way. However, only a more systematic study of the formation efficiency of XRBs in the higher metallicity LMC galaxy will show how it truly depends on the metallicity.

\section{Conclusions} \label{sec:Conclusions}

We have investigated the formation efficiency of HMXBs in the low SMC metallicity {\it for the first time} as a function of the age of their parent stellar population. We have used the different formation efficiency indicators N(HMXBs)/SFR, N(HMXBs)/N(OBs), and N(HMXBs)/M$\star$, all as a function of the age of the major associated SF burst. In all cases, we find an increase in the formation efficiency up to an age of $\sim$40--60\,Myr, and a gradual decrease thereafter. The peak formation efficiency N(HMXB)/SFR is estimated as (49 $\pm$ 14) $[10^{-5}~{\rm M_{\odot}/yr}]^{-1}$, in good agreement with previous estimates of the average formation efficiency in the broad $\sim$20--60\,Myr age range. This peak in the formation efficiency of the SMC HMXBs (8$\times$ and 4$\times$ higher than at earlier ($\sim$10\,Myr), and later epochs ($\sim$260\,Myr), respectively) is in excellent agreement with previous studies that have examined it on Be stars in both the Magellanic Clouds and the Milky Way.

\section*{Acknowledgements}

We thank K. D. Kuntz for useful comments that have improved the quality of the paper. VA acknowledges financial support from NASA/Chandra grants GO3-14051X, AR4-15003X, NNX15AR30G, NASA/ADAP grant NNX10AH47G, and the Texas Tech President's Office. AZ acknowledges financial support from NASA/ADAP grant NNX12AN05G and funding from the European Research Council under the European Union's Seventh Framework Programme (FP/2007-2013)/ERC Grant Agreement n.~617001. This project has also received funding from the European Union's Horizon 2020 research and innovation programme  under the Marie Sklodowska-Curie RISE action,  grant agreement No 691164 (ASTROSTAT). JJD, TJG, and PPP were funded by NASA contract NAS8-03060 to the {\it Chandra X-ray Center}. PFW acknowledges financial support from the NSF through grant AST-1714281. The OGLE project has received funding from the National Science Centre, Poland, grant MAESTRO 2014/14/A/ST9/00121 to AU. We thank the CXC Director, Belinda Wilkes, for advice and support, and for funding the publication of this work. This research has made use of NASA's Astrophysics Data
System and the Tool for OPerations on Catalogues And Tables (TOPCAT)
software package \citep{2005ASPC..347...29T}.

\end{document}

%% file: definitions.tex
\newcommand{\asca}{{\it ASCA }}

\newcommand{\rxte}{{\it RXTE }}
\newcommand{\integral}{{\it INTEGRAL }}

\newcommand{\einstein}{{\it Einstein }}  % ApJ uses Italics for 'Einstein' Satellite 

\newcommand{\funit}{~erg cm$^{-2}$ s$^{-1}$} 
\newcommand{\ergs}{~erg s$^{-1}$}             
\newcommand{\msun}{~M$_{\odot}$\,}
\newcommand{\msunsimple}{M$_{\odot}$}

\newcommand{\zsun}{Z$_{\odot}$}

\newcommand{\rosat}{{\it ROSAT }}

\newcommand{\chandra}{{\it Chandra~}}

\newcommand{\xmm}{{\it XMM-Newton~}}

\newcommand{\Ha}{H$\alpha$~}

%% file: HMXBs_FormEff.bbl
\begin{thebibliography}{}

%\bibitem[Anders \& Grevesse(1989)]{1989GeCoA..53..197A} Anders, E., \& Grevesse, N.\ 1989, \gca, 53, 197%Abundances of the elements - Meteoritic and solar


\bibitem[Andrews et al.(2018)]{2018ApJS..237....1A} Andrews, J.~J., Zezas, A., \& Fragos, T.\ 2018, \apjs, 237, 1%dart_board: Binary Population Synthesis with Markov Chain Monte Carlo



\bibitem[Antoniou et al.(2009b)]{2009ApJ...707.1080A} Antoniou, V., Hatzidimitriou, D., Zezas, A., \& Reig, P.\ 2009b, \apj, 707, 1080%Optical Spectroscopy of 20 Be/X-ray Binaries in the Small Magellanic Cloud

\bibitem[Antoniou \& Zezas(2016)]{2016MNRAS.459..528A} Antoniou, V., \& Zezas, A.\ 2016, \mnras, 459, 528 

\bibitem[Antoniou et al.(2010)]{2010ApJ...716L.140A} Antoniou, V., Zezas, A., Hatzidimitriou, D., \& Kalogera, V.\ 2010, \apjl, 716, L140 

\bibitem[Antoniou et al.(2009a)]{2009ApJ...697.1695A} Antoniou, V., Zezas, A., Hatzidimitriou, D., \& McDowell, J.~C.\ 2009a, \apj, 697, 1695%The Chandra Survey of the Small Magellanic Cloud "Bar". II. Optical Counterparts of X-ray Sources


%\bibitem[Asplund et al.(2009)]{2009ARA&A..47..481A} Asplund, M., Grevesse, N., Sauval, A.~J., \& Scott, P.\ 2009, \araa, 47, 481%The Chemical Composition of the Sun

\bibitem[Badenes et al.(2015)]{2015ApJ...804L..25B} Badenes, C., Maoz, D., \& Ciardullo, R.\ 2015, \apjl, 804, L25%The Progenitors and Lifetimes of Planetary Nebulae


\bibitem[Belczynski et al.(2008)]{2008ApJS..174..223B} Belczynski, K., Kalogera, V., Rasio, F.~A., et al.\ 2008, \apjs, 174, 223-260%Compact Object Modeling with the StarTrack Population Synthesis Code

%\bibitem[Binder et al.(2015)]{2015AJ....150...94B} Binder, B., Williams, B.~F., Eracleous, M., et al.\ 2015, \aj, 150, 94%The Chandra Local Volume Survey. I. The X-Ray Point Source Populations of NGC 55, NGC 2403, and NGC 4214 

\bibitem[Bonanos et al.(2009)]{2009AJ....138.1003B} Bonanos, A.~Z., Massa, D.~L., Sewilo, M., et al.\ 2009, \aj, 138, 1003%Spitzer SAGE Infrared Photometry of Massive Stars in the Large Magellanic Cloud

\bibitem[Bonanos et al.(2010)]{2010AJ....140..416B} Bonanos, A.~Z., Lennon, D.~J., K{\"o}hlinger, F., et al.\ 2010, \aj, 140, 416%Spitzer SAGE-SMC Infrared Photometry of Massive Stars in the Small Magellanic Cloud

\bibitem[Bressan et al.(2012)]{2012MNRAS.427..127B} Bressan, A., Marigo, P., Girardi, L., et al.\ 2012, \mnras, 427, 127%PARSEC: stellar tracks and isochrones with the PAdova and TRieste Stellar Evolution Code

\bibitem[\protect\citeauthoryear{Broos et al.}{2010}]{Broos...et...al...2010} Broos P.~S., Townsley L.~K., Feigelson E.~D., Getman K.~V., Bauer F.~E., Garmire G.~P., 2010, ApJ, 714, 1582 
%\bibitem[\protect\citeauthoryear{Broos et al.}{2010}]{2010ApJ...714.1582B} Broos P.~S., Townsley L.~K., Feigelson E.~D., Getman K.~V., Bauer F.~E., Garmire G.~P., 2010, ApJ, 714, 1582%Innovations in the Analysis of Chandra-ACIS Observations 

\bibitem[\protect\citeauthoryear{Broos et al.}{2012}]{2012ascl.soft03001B} Broos P., Townsley L., Getman K., Bauer F., 2012, ascl.soft, 1203.001
%\bibitem[\protect\citeauthoryear{Broos et al.}{2012}]{2012ascl.soft03001B} Broos P., Townsley L., Getman K., Bauer F., 2012, ascl.soft, 1203.001%AE: ACIS Extract 

\bibitem[Cardelli et al.(1989)]{1989ApJ...345..245C} Cardelli, J.~A., Clayton, G.~C., \& Mathis, J.~S.\ 1989, \apj, 345, 245%The relationship between infrared, optical, and ultraviolet extinction

\bibitem[Coe \& Kirk(2015)]{2015MNRAS.452..969C} Coe, M.~J., \& Kirk, J.\ 2015, \mnras, 452, 969%Catalogue of Be/X-ray binary systems in the Small Magellanic Cloud: X-ray, optical and IR properties

\bibitem[Corbet et al.(2009)]{2009IAUS..256..361C} Corbet, R.~H.~D., Coe, M.~J., McGowan, K.~E., et al.\ 2009, The Magellanic System: Stars, Gas, and Galaxies, 256, 361%Properties of X-ray binaries in the Magellanic Clouds from RXTE and Chandra observations

\bibitem[de Grijs \& Bono(2015)]{2015AJ....149..179D} de Grijs, R., \& Bono, G.\ 2015, \aj, 149, 179%Clustering of Local Group Distances: Publication Bias or Correlated Measurements? III. The Small Magellanic Cloud

\bibitem[Dray(2006)]{2006MNRAS.370.2079D} Dray, L.~M.\ 2006, \mnras, 370, 2079%On the metallicity dependence of high-mass X-ray binaries

%\bibitem[\protect\citeauthoryear{Evans et al.}{2010}]{Evans...et...al...2010}  Evans I.~N., et al., 2010, ApJS, 189, 37
%\bibitem[\protect\citeauthoryear{Evans et al.}{2010}]{2010ApJS..189...37E}  Evans I.~N., et al., 2010, ApJS, 189, 37%The Chandra Source Catalog

\bibitem[Fragos et al.(2013)]{2013ApJ...764...41F} Fragos, T., Lehmer, B., Tremmel, M., et al.\ 2013, \apj, 764, 41%X-Ray Binary Evolution Across Cosmic Time

\bibitem[Gordon et al.(2003)]{2003ApJ...594..279G} Gordon, K.~D., Clayton, G.~C., Misselt, K.~A., Landolt, A.~U., \& Wolff, M.~J.\ 2003, \apj, 594, 279%A Quantitative Comparison of the Small Magellanic Cloud, Large Magellanic Cloud, and Milky Way Ultraviolet to Near-Infrared Extinction Curves

\bibitem[Grimm et al.(2003)]{2003MNRAS.339..793G} Grimm, H.-J., Gilfanov, M., \& Sunyaev, R.\ 2003, \mnras, 339, 793%High-mass X-ray binaries as a star formation rate indicator in distant galaxies

\bibitem[Haberl et al.(2000)]{2000A&AS..142...41H} Haberl, F., Filipovi{\'c}, M.~D., Pietsch, W., \& Kahabka, P.\ 2000, \aaps, 142, 41%A ROSAT PSPC catalogue of X-ray sources in the SMC region

\bibitem[Haberl et al.(2012)]{2012A&A...545A.128H} Haberl, F., Sturm, R., Ballet, J., et al.\ 2012, \aap, 545, A128%The XMM-Newton survey of the Small Magellanic Cloud

\bibitem[Haberl \& Sturm(2016)]{2016A&A...586A..81H} Haberl, F., \& Sturm, R.\ 2016, \aap, 586, A81%High-mass X-ray binaries in the Small Magellanic Cloud 


%\bibitem[Harries et al.(2003)]{2003MNRAS.339..157H} Harries, T.~J., Hilditch, R.~W., \& Howarth, I.~D.\ 2003, \mnras, 339, 157%Ten eclipsing binaries in the Small Magellanic Cloud: fundamental parameters and Cloud distance

\bibitem[\protect\citeauthoryear{Harris \& Zaritsky}{2004}]{2004AJ....127.1531H} Harris J., Zaritsky D., 2004, AJ, 127, 1531%The Star Formation History of the Small Magellanic Cloud


\bibitem[Harrison et al.(2013)]{2013ApJ...770..103H}  Harrison, F.~A., Craig, W.~W., Christensen, F.~E., et al.\ 2013, \apj, 770, 103%The Nuclear Spectroscopic Telescope Array (NuSTAR) High-energy X-Ray Mission

\bibitem[Hong et al.(2005)]{2005ApJ...635..907H} Hong, J., van den Berg, M., Schlegel, E.~M., et al.\ 2005, \apj, 635, 907%X-Ray Processing of ChaMPlane Fields: Methods and Initial Results for Selected Anti-Galactic Center Fields

\bibitem[Hong et al.(2016)]{2016ApJ...826....4H} Hong, J., Antoniou, V., Zezas, A., et al.\ 2016, \apj, 826, 4%SXP 214: An X-Ray Pulsar in the Small Magellanic Cloud, Crossing the Circumstellar Disk of the Companion

%\bibitem[Hong et al.(2017)]{Jaesub...PulsarsCatalog} Hong, J., et al.\ 2017, \apj, submitted
\bibitem[Hong et al.(2017)]{2017ApJ...847...26H} Hong, J., Antoniou, V., Zezas, A., et al.\ 2017, \apj, 847, 26 
%Deep Chandra Survey of the Small Magellanic Cloud. II. Timing Analysis of X-Ray Pulsars

%\bibitem[\protect\citeauthoryear{Illarionov \& Sunyaev}{1975}]{1975A&A....39..185I} Illarionov A.~F., Sunyaev R.~A., 1975, A\&A, 39, 185%Why the Number of Galactic X-ray Stars Is so Small?

\bibitem[Iqbal \& Keller(2013)]{2013MNRAS.435.3103I} Iqbal, S., \& Keller, S.~C.\ 2013, \mnras, 435, 3103%A study of Be stars in the Magellanic Clouds

\bibitem[Laycock et al.(2005)]{2005ApJS..161...96L} Laycock, S., Corbet, R.~H.~D., Coe, M.~J., et al.\ 2005, \apjs, 161, 96%Long-Term Behavior of X-Ray Pulsars in the Small Magellanic Cloud

\bibitem[\protect\citeauthoryear{Laycock et al.}{2010}]{2010ApJ...716.1217L} Laycock S., Zezas A., Hong J., Drake J.~J., Antoniou V., 2010, ApJ, 716, 1217%Exploring the Small Magellanic Cloud to the Faintest X-ray Fluxes: Source Catalog, Timing, and Spectral Analysis

\bibitem[Lehmer et al.(2010)]{2010ApJ...724..559L} Lehmer, B.~D., Alexander, D.~M., Bauer, F.~E., et al.\ 2010, \apj, 724, 559%A Chandra Perspective on Galaxy-wide X-ray Binary Emission and its Correlation with Star Formation Rate and Stellar Mass: New Results from Luminous Infrared Galaxies

\bibitem[Lehmer et al.(2016)]{2016ApJ...825....7L} Lehmer, B.~D., Basu-Zych, A.~R., Mineo, S., et al.\ 2016, \apj, 825, 7%The Evolution of Normal Galaxy X-Ray Emission through Cosmic History: Constraints from the 6 MS Chandra Deep Field-South

\bibitem[Linden et al.(2010)]{2010ApJ...725.1984L} Linden, T., Kalogera, V., Sepinsky, J.~F., et al.\ 2010, \apj, 725, 1984%The Effect of Starburst Metallicity on Bright X-ray Binary Formation Pathways 

\bibitem[Luck et al.(1998)]{1998AJ....115..605L} Luck, R.~E., Moffett, T.~J., Barnes, T.~G., III, \& Gieren, W.~P.\ 1998, \aj, 115, 605%Magellanic Cloud Cepheids - Abundances

\bibitem[Maravelias et al.(2014)]{2014MNRAS.438.2005M} Maravelias, G., Zezas, A., Antoniou, V., \& Hatzidimitriou, D.\ 2014, \mnras, 438, 2005%Optical spectra of five new Be/X-ray binaries in the Small Magellanic Cloud and the link of the supergiant B[e] star LHA 115-S 18 with an X-ray source

\bibitem[Martayan et al.(2006)]{2006A&A...452..273M} Martayan, C., Fr{\'e}mat, Y., Hubert, A.-M., et al.\ 2006, \aap, 452, 273%Effects of metallicity, star-formation conditions, and evolution in B and Be stars. I. Large Magellanic Cloud, field of NGC 2004


\bibitem[Martayan et al.(2007)]{2007A&A...472..577M} Martayan, C., Floquet, M., Hubert, A.~M., et al.\ 2007, \aap, 472, 577%Be stars and binaries in the field of the SMC open cluster NGC 330 with VLT-FLAMES


\bibitem[Martayan et al.(2007)]{2007A&A...462..683M} Martayan, C., Fr{\'e}mat, Y., Hubert, A.-M., et al.\ 2007, \aap, 462, 683%Effects of metallicity, star-formation conditions, and evolution in B and Be stars. II. Small Magellanic Cloud, field of NGC?330

\bibitem[McBride et al.(2008)]{2008MNRAS.388.1198M} McBride, V.~A., Coe, M.~J., Negueruela, I., Schurch, M.~P.~E., \& McGowan, K.~E.\ 2008, \mnras, 388, 1198%Spectral distribution of Be/X-ray binaries in the Small Magellanic Cloud


\bibitem[McGowan et al.(2008)]{2008MNRAS.383..330M} McGowan, K.~E., Coe, M.~J., Schurch, M.~P.~E., et al.\ 2008, \mnras, 383, 330%The Chandra Small Magellanic Cloud Wing Survey - the search for X-ray binaries 


\bibitem[McSwain \& Gies(2005)]{2005ApJS..161..118M} McSwain, M.~V., \& Gies, D.~R.\ 2005, \apjs, 161, 118%The Evolutionary Status of Be Stars: Results from a Photometric Study of Southern Open Clusters

\bibitem[Mineo et al.(2012)]{2012MNRAS.419.2095M} Mineo, S., Gilfanov, M., \& Sunyaev, R.\ 2012, \mnras, 419, 2095%X-ray emission from star-forming galaxies - I. High-mass X-ray binaries 


\bibitem[\protect\citeauthoryear{Naz{\'e} et al.}{2002}]{2002ApJ...580..225N} Naz{\'e} Y., Hartwell J.~M., Stevens I.~R., Corcoran M.~F., Chu Y.-H., Koenigsberger G., Moffat A.~F.~J., Niemela V.~S., 2002, ApJ, 580, 225%An X-Ray Investigation of the NGC 346 Field in the Small Magellanic Cloud I. The Luminous Blue Variable HD 5980 and the NGC 346 Cluster


\bibitem[Porter \& Rivinius(2003)]{2003PASP..115.1153P} Porter, J.~M., \& Rivinius, T.\ 2003, \pasp, 115, 1153%Classical Be Stars

\bibitem[Sasaki et al.(2000)]{2000A&AS..147...75S} Sasaki, M., Haberl, F., \& Pietsch, W.\ 2000, \aaps, 147, 75%ROSAT HRI catalogue of X-ray sources in the SMC region

\bibitem[Seward \& Mitchell(1981)]{1981ApJ...243..736S} Seward, F.~D., \& Mitchell, M.\ 1981, \apj, 243, 736%X-ray survey of the Small Magellanic Cloud

\bibitem[Shtykovskiy \& Gilfanov(2005)]{2005A&A...431..597S} Shtykovskiy, P., \& Gilfanov, M.\ 2005, \aap, 431, 597%High mass X-ray binaries in the LMC: Dependence on the stellar population age and the ``propeller'' effect


\bibitem[Shtykovskiy \& Gilfanov(2007)]{2007AstL...33..437S} Shtykovskiy, P.~E., \& Gilfanov, M.~R.\ 2007, Astronomy Letters, 33, 437%High-mass X-ray binaries and recent star formation history of the Small Magellanic Cloud

\bibitem[Sturm et al.(2013)]{2013A&A...558A...3S} Sturm, R., Haberl, F., Pietsch, W., et al.\ 2013, \aap, 558, A3%The XMM-Newton survey of the Small Magellanic Cloud: The X-ray point-source catalogue

\bibitem[Taylor(2005)]{2005ASPC..347...29T} Taylor, M.~B.\ 2005, Astronomical Data Analysis Software and Systems XIV, 347, 29%TOPCAT & STIL: Starlink Table/VOTable Processing Software

\bibitem[Udalski et al.(1999)]{1999AcA....49..437U} Udalski, A., Soszynski, I., Szymanski, M., et al.\ 1999, \actaa, 49, 437%The Optical Gravitational Lensing Experiment. Cepheids in the Magellanic Clouds. V. Catalog of Cepheids from the Small Magellanic Cloud

\bibitem[Udalski et al.(2008)]{2008AcA....58..329U} Udalski, A., Soszy{\'n}ski, I., Szyma{\'n}ski, M.~K., et al.\ 2008, \actaa, 58, 329%The Optical Gravitational Lensing Experiment. OGLE-III Photometric Maps of the Small Magellanic Cloud

\bibitem[Walter et al.(2015)]{2015A&ARv..23....2W} Walter, R., Lutovinov, A.~A., Bozzo, E., \& Tsygankov, S.~S.\ 2015, \aapr, 23, 2%High-mass X-ray binaries in the Milky Way. A closer look with INTEGRAL

\bibitem[Williams et al.(2013)]{2013ApJ...772...12W} Williams, B.~F., Binder, B.~A., Dalcanton, J.~J., Eracleous, M., \& Dolphin, A.\ 2013, \apj, 772, 12%The Ages of High-mass X-Ray Binaries in NGC 2403 and NGC 300

\bibitem[Yokogawa et al.(2003)]{2003PASJ...55..161Y} Yokogawa, J., Imanishi, K., Tsujimoto, M., Koyama, K., \& Nishiuchi, M.\ 2003, \pasj, 55, 161%Review of Discrete X-Ray Sources in the Small Magellanic Cloud: Summary of the ASCA Results and Implication on the Recent Star-Forming Activity


\end{thebibliography}
